\documentclass[preprint, 5p, onecolumn, sort&compress]{elsarticle}
\usepackage{amsmath}
\usepackage{amssymb}
\usepackage{graphicx}
\usepackage{subdepth}
\usepackage[italic]{hepparticles}
\usepackage[italic]{hepnames}
\usepackage{natbib}
\usepackage[usenames,dvipsnames]{color}

\usepackage{hyperref}
\journal{J. High Energy Astrophys. 32 (2021) 1-5, \url{https://doi.org/10.1016/j.jheap.2021.07.001}}

\begin{document}

    \begin{frontmatter}

\newcommand*{\PKU}{School of Physics and State Key Laboratory of Nuclear Physics and Technology, Peking University, Beijing 100871, China}
\newcommand*{\CHEP}{Center for High Energy Physics, Peking University, Beijing 100871, China}
\newcommand*{\CIC}{Collaborative Innovation Center of Quantum Matter, Beijing, China}

    \title{Threshold anomalies of ultra-high energy cosmic photons due to Lorentz invariance violation}

    \author[a]{Hao Li}

    \author[a,b,c]{Bo-Qiang Ma\corref{cor1}}

    \address[a]{\PKU}
\address[b]{\CHEP}
\address[c]{\CIC}
\cortext[cor1]{Corresponding author \ead{mabq@pku.edu.cn}}


    \begin{abstract}
From special relativity, photon annihilation process \HepProcess{{\Pgg}{\Pgg}{\to}{\Pep}{\Pem}} prevents cosmic photons with energies above a threshold to propagate a long distance in cosmic space due to their annihilation with low energy cosmic background photons. However, modifications of the photon dispersion relation from Lorentz invariance violation~(LIV) can cause novel phenomena beyond special relativity to happen. In this paper, we point out that these phenomena include optical transparency, threshold reduction and reappearance of ultra-high energy photons in cosmic space. The recent observation of near and above PeV photon events by the LHAASO Collaboration reveals the necessity to consider the threshold anomalies.  Future observations of above threshold photons from extragalactic sources can testify LIV properties of photons.
    \end{abstract}


    \end{frontmatter}


    Symmetries play important roles in physics, and the Lorentz symmetry as well as consequent Lorentz invariance, as a fundamental principle in special relativity, is considered as a profound building block of modern physics. Lorentz invariance has been confirmed by almost all experiments to high precision. Regardless of these experimental constraints, certain quantum gravity~(QG) theories, such as stringy theories~\cite{Amelino-Camelia1997a, Ellis1999, Ellis2000,Ellis2008, Li2009, Li:2021pre, LI2021104380}, loop quantum gravity~\cite{Gambini1999}, and special-relativity-like theories such as doubly special relativity~\cite{AMELINO-CAMELIA2002a,AMELINO-CAMELIA2002b, Magueijo2002}, do predict the violation of Lorentz invariance~\cite{Mattingly2005, Jacobson2006, Liberati2009, Amelino-Camelia2013}. In order to reveal the Lorentz invariance violation~(LIV) properties of particles, we can utilize LIV induced dispersion relations which are model dependent. However no matter what the underlying theories are, we can always adopt a model independent form to describe the modified dispersion relation\footnote{Provided that rotation invariance is preserved and the energy \(E\) of the particle satisfies \(E \ll E_{\text{Planck}}\simeq 1.22\times 10^{19}~\text{GeV}\).} as~\cite{Xiao2009, Shao2010b}:
    \begin{equation}
        E^2(p) = p^2 + m^2 - \eta p^n,
        \label{generalrelation}
    \end{equation}
    where \(m\) is the mass of the particle, \(p\) is the magnitude of the momentum \(\vec{p}\), \(\eta \) is a particle species dependent parameter which is highly suppressed\footnote{Generally speaking, it is expected to be suppressed by the Planck scale, see, e.g., Ref.~\cite{AMELINO-CAMELIA2002a}.} and could be positive, negative or just zero, and \(n = 3, 4, \cdots \) where we have dropped higher order contributions since they are much smaller.

    In most cases, LIV effects are so tiny that there are only some special experiments and observations to constrain the LIV parameters. For example, it is widely agreed and suggested that considering the smallness of the LIV effects, high energy photons from distant objects such as gamma-ray bursts~(GRBs) could provide a unique opportunity to reveal LIV properties of photons~\cite{Amelino-Camelia1998, Ellis2006, Jacob2008}. Intensive researches have been performed on time of light flight measurements and these researches require the knowledge of the sources such as the redshifts. This kind of work revealed some intriguing LIV properties of photons and we discuss it again at the end of this article.

    The purpose of this paper is to focus on a different method to reveal LIV properties of ultra-high energy~(UHE) cosmic photons through the novel threshold anomalies of photon annihilation process \HepProcess{{\Pgg}{\Pgg}{\to}{\Pep}{\Pem}}. From special relativity, the annihilation of two photons into an electron-positron pair occurs above certain~(lower) threshold\footnote{By threshold here, we mean that after fixing the energy of one photon, the smallest~(lower threshold) or the largest~(upper threshold) energy of the other photon to allow the process to happen.} and such annihilation prevents UHE photons from propagating a long distance in cosmic space because of the attenuation by the cosmic background lights such as cosmic microwave background~(CMB), interstellar radiation fields~(ISRF) and extragalactic background light~(EBL). However deviation from Lorentz invariance can modify this threshold behavior drastically and even a very simple model of the modified dispersion relation can allow novel physical phenomena beyond special relativity to occur.

\begin{figure}
        \centering
        \includegraphics[scale=1]{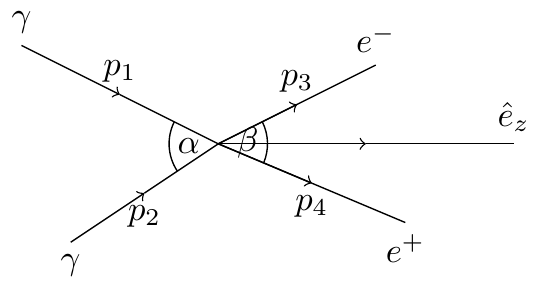}
        \caption{\label{geom} The 
        diagram of photon annihilation into an electron-positron pair.}
    \end{figure}

    We now briefly review the standard threshold from special relativity and discuss in detail the modified threshold effects.
    The 
    diagram of \HepProcess{{\Pgg}{\Pgg}{\to}{\Pep}{\Pem}} is shown in Fig.~\ref{geom}, where the angles between the two incoming photons and outgoing electron/positron are \(\alpha \) and \(\beta \) respectively. Because we merely focus on thresholds, we take the photon with 4-momentum \(p_2\) as the fixed one, and assume that its energy \(\varepsilon_b\) as well as its momentum magnitude \({\left\lvert\vec{p_2}\right\rvert}\) is much smaller than that of the other photon with 4-momentum \(p_1\).

    From special relativity, the lower threshold of this process occurs when \(\alpha = \pi \), \(\beta = 0\) and the electron~(positron) carries half of the incoming energy and momentum\footnote{One can verify this by considering Lorentz transformation from the center-of-mass reference frame to the experimental frame.} and there is no upper threshold. It is straightforward to calculate the threshold since we already determined the threshold configuration:
    \[p_1=\left(E, 0, 0, E\right),\]
    \[p_2 = \left(\varepsilon_b, 0, 0, -\varepsilon_b\right),\]
    \[p_3 = p_4 = \left(\frac{E+\varepsilon_b}{2}, 0, 0, \frac{E-\varepsilon_b}{2}\right).\]
    Utilizing the on-shell condition for the electron, we have the identity,
    \begin{equation}
        m_e^2 \equiv p_3^2 = {\left(\frac{E+\varepsilon_b}{2}\right)}^2 - {\left(\frac{E-\varepsilon_b}{2}\right)}^2,
    \end{equation}
    which reads \(E = \dfrac{m_e^2}{\varepsilon_b}\). Therefore we obtain the threshold condition for photon annihilation into an electron-positron pair from special relativity:
    \begin{equation}
        E \ge E_{\text{th}} = \frac{m_e^2}{\varepsilon_b}.\label{smthres}
    \end{equation}
    Once the energy of a photon exceeds \(m^2_e/\varepsilon_b\), there are certain configurations allowing this reaction to occur and causing photon attenuation by the low energy photon with 4-momentum \(p_2\). Otherwise the reaction is kinematically forbidden.

    However, slight modifications from LIV can drastically spoil this simple threshold behavior and more interesting phenomena beyond special relativity can emerge. Before going on, we further assume that the rotation symmetry as well as energy-momentum conservation is preserved. Therefore for a photon, we can write down the modified dispersion relation according to Eq.~\eqref{generalrelation},
    \begin{equation}
        \omega^2(k) = k^2 - \xi k^n, \label{photondis}
    \end{equation}
    where \(\xi \) is the LIV parameter which could be positive~(subluminal), negative~(superluminal) or zero. Furthermore, we assume that electrons and positrons do not have any LIV modification for simplification, and observations as well as some theory models strongly support this assumption~\cite{Ellis2008, Li2009, Maccione2007, Ellis2009}. Again we adopt the same threshold configuration in Fig.~\ref{geom} as in the special relativity case~\cite{Mattingly2003}, while now there might be an upper threshold~\cite{Mattingly2003, Jacobson2003} if the LIV parameter \(\xi \) takes values in a specific interval. The possibility of the existence of an upper threshold for two photon annihilation into an electron-positron pair was first pointed out by
    Klu\'zniak~\cite{KLUZNIAK1999117}. We need to modify \(p_1\) which we introduced in the special relativity case to \(p_1 = (\omega(k),0,0,k)\), where \(\omega(k)\) is defined in Eq.~\eqref{photondis}, and keep other momenta unchanged\footnote{It is not necessary to modify \(p_2\), because \(\varepsilon_b \) is so small that contributions from LIV are negligible.}. Since we assume the conservation of energy-momentum, utilizing the on-shell condition for the outgoing electron, after some manipulations like what we do in the previous paragraph, we have,
    \begin{align}
        m_e^2 &= {\left(\frac{\omega + \varepsilon_b}{2}\right)}^2 - {\left(\frac{k-\varepsilon_b}{2}\right)}^2 \notag \\
              &= \frac{\omega^2-k^2+2\varepsilon_b\left(\omega+k\right)}{4}\notag \\
              &= \frac{-\xi k^n + 2\varepsilon_b\left(2k - \frac{1}{2}\xi k^{n-1}\right)}{4} + O\left(\xi^2\right)\notag \\
              &= \frac{-\xi k^n + 4k\varepsilon_b}{4} + O\left(\xi^2\right)+O\left(\xi\epsilon_b\right),
    \end{align}
    or equivalently,
    \begin{equation}
        \xi = f(k) = \frac{4\varepsilon_b}{k^{n-1}} - \frac{4 m_e^2}{k^n}, \text{ for }k>0,\label{maineq}
    \end{equation}
    where we define a function \(f(k)\). Hereafter we consider a simplified situation where we can take \(n\) in Eq.~\eqref{maineq} as 3 which corresponds to the well studied case in phenomenological studies. To study the threshold behaviors now is equal to study the solutions of Eq.~\eqref{maineq}. Furthermore, one can finger out easily how many and where the solutions of Eq.~\eqref{maineq} are by investigating the relation between the value of \(\xi \) and the function \(f(k)\). Hence it is quite helpful to study the properties of the function \(f(k) \). We restrict ourselves to \(k>0\) since \(k<0\) is meaningless. The useful properties of \(f(k)\) are as follows:
    \begin{enumerate}
        \item[*] There is only one zero point, \(k_0 = \dfrac{m_e^2}{\varepsilon_b}\), which is just the threshold derived in the special relativity case;
        \item[*] The function tends to zero at \(k = +\infty \) and tends to \(-\infty \) at \(k=0\);
        \item[*] There is a maximum at the critical point \(k_c=\dfrac{3m_e^2}{2\varepsilon_b}\) with \(\xi_c=\max{f(k)} = f(k_c) = \dfrac{16 \varepsilon_b^3}{27 m_e^4}\).
    \end{enumerate}

\begin{figure}
        \centering
        \includegraphics[scale=0.5]{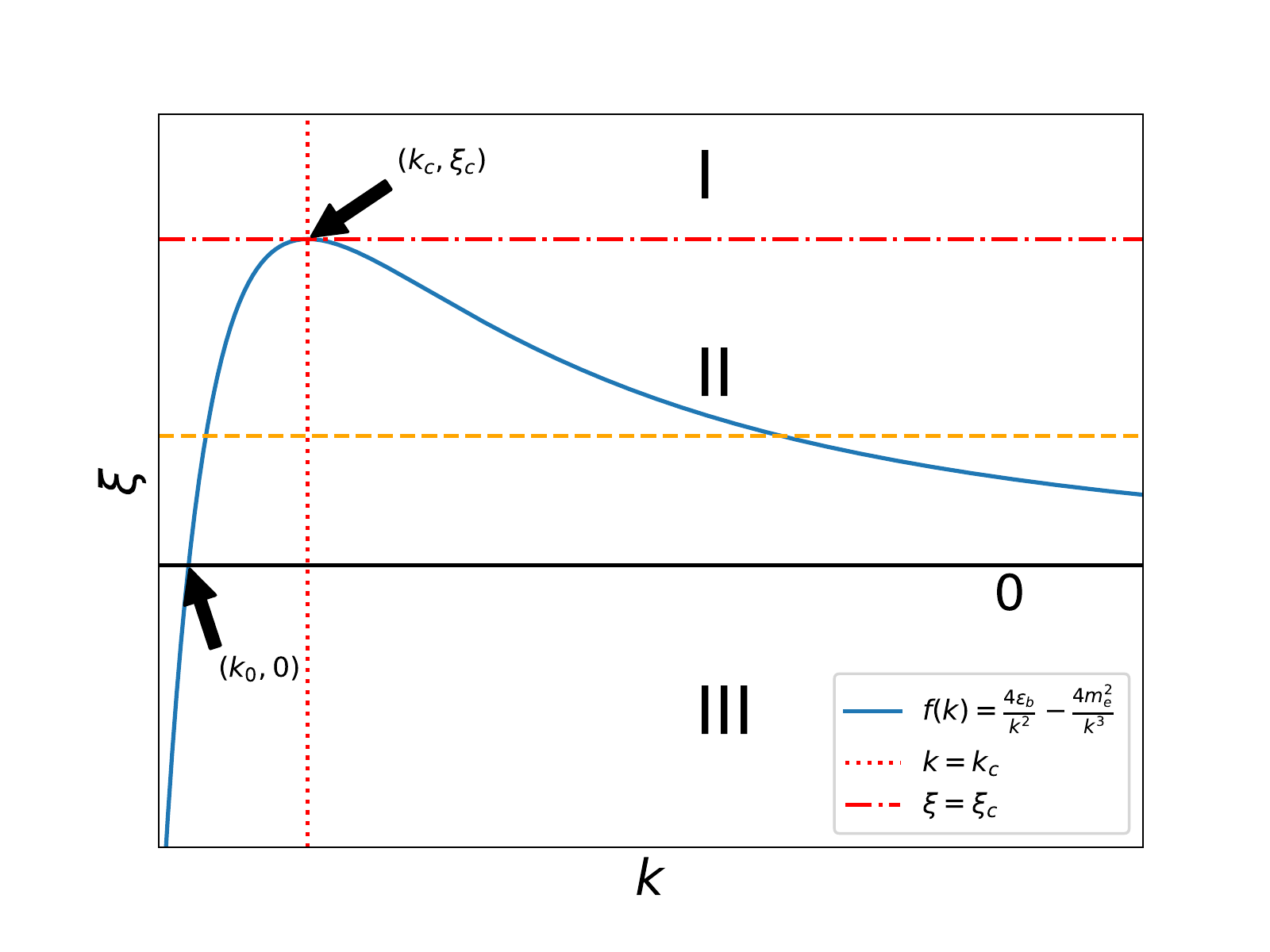}
        \caption{\label{func} Graphical representation of the relation between \(\xi \) and \(f(k)\). The horizontal orange dashed line indicates an example of {\bfseries Case II} with fixed \(\xi \) in {\bfseries Region II}.}
    \end{figure}

    With these preparations, we represent the LIV parameter \(\xi \) and the image of \(f(k)\) schematically in Fig.~\ref{func}, which turns out to be quite inspiring. We divide the codomain of the function \(f(k)\) into three regions: {\bfseries Region I}~(\(f(k)>\xi_c \)), {\bfseries Region II}~(\(0<f(k)<\xi_c\)) and {\bfseries Region III}~(\(f(k)<0\)) which correspond to three different threshold behaviors. What we can learn from Fig.~\ref{func} are as follows:
    \begin{description}
        \item[Case I~(optical transparency)] If \(\xi \) falls into {\bfseries Region I},
        there is no solution for Eq.~\eqref{maineq}, which means that even a lower threshold does not exist. This case corresponds to subluminal photons, and hence we conclude that subluminal photons
        with $\xi> \xi_c$
        cannot be absorbed by photons with energy \(\varepsilon_b\).
        \item[Case II~(reappearance of UHE photons)] If \(\xi \) falls into {\bfseries Region II},
        there are two distinct solutions for Eq.~\eqref{maineq}, which are denoted by \(k_<\) and \(k_>\) for the smaller and larger one respectively\footnote{From Fig.~\ref{func}, it is clear that \(0<k_<<k_c<k_>\).}. It is obvious that \(k_<\) is the lower threshold, while there is an extra solution \(k_>\). With the help of the theorems in Ref.~\cite{Mattingly2003}, we eventually identify \(k_>\) as an upper threshold. In conclusion, in this case only photons with energies between \(\omega(k_<)\) and \(\omega(k_>)\) are kinematically allowed to be absorbed by the photons with energy \(\varepsilon_b\) to create an electron-positron pair. Therefore we obtain an intriguing picture: the background light with energy \(\varepsilon_b\) is optical transparent to photons with energies below \(\omega(k_<)\) as usual, and it is again transparent to photons with energies above \(\omega(k_>)\). Finally, it is worth noting that when \(\xi \) tends to zero, \(k_> \) tends to \(+\infty \). That is to say we return to the special relativity case, which is expected since any quantum gravity theory must have standard model as its low energy effective field theory.
        \item[Case III~(threshold reduction)] If \(\xi \) falls into {\bfseries Region III},
        there is only one solution for Eq.~\eqref{maineq}, which means there is a lower threshold. It is clear that the solution is smaller than \(k_0\), therefore the threshold behavior here is almost the same as the case in special relativity, except for the reduction of the lower threshold.
    \end{description}

    Now it is convenient to discuss what we can learn from the threshold anomalies discussed above.
    When photons are superluminal, which corresponds to {\bfseries Case III}, there is a threshold reduction of the photon annihilation process but other processes such as {\Pgg} decay can set stronger constraints on \(\xi \).
    We then concentrate on {\bfseries Case I} and {\bfseries Case II}\@. To be more specific, we assume that the photon with 4-momentum \(p_2\) is from CMB, and its mean energy \(\varepsilon_b\) is about \(6.35\times 10^{-4}~\text{eV}\). As a result, \(k_c=\frac{3}{2}k_0\simeq 617~\text{TeV}\). Due to the absence of a well-accepted quantum gravity theory, we expect our discussion is only valid for photons with energies much less than \(E_\text{Planck}\), so that the discussion is still reasonable when the photon energy is around several EeVs. From special relativity, if the energy of a photon exceeds \(k_0\simeq 411~\text{TeV}\), there is an obvious attenuation of this photon by CMB\@. However, novel phenomena beyond special relativity emerge if LIV exists. If we observe excesses of the number of photons with energies above \(k_0\simeq 411~\text{TeV}\) and associate them with distant extragalactic sources as {\bfseries Case I} and {\bfseries Case II} both indicate, we can attribute these observations as the support for the photon LIV properties predicted above. We suggest ground-based cosmic-ray observatories such as the Large High Altitude Air Shower Observatory~(LHAASO)~\cite{Cao2010, Cao2019}, which is one of the most sensitive gamma-ray detector arrays currently operating at TeV and PeV energies, as an ideal platform to perform such observations.

    From the speed-energy relation \(v(E)=\dfrac{\partial \omega(k)}{\partial k}\), we can derive the modified speed of light \(v(E)\) from Eq.~\eqref{photondis},
    \begin{equation}
        v(E) = c\left(1-\frac{n-1}{2}\xi E^{n-2}\right) + O\left(\xi^2\right)\label{speedoflight},
    \end{equation}
    where we restore the light speed constant \(c\) for low energy photons. In fact, intensive studies have been performed on the \(n=3\) case of Eq.~\eqref{speedoflight} based on time of light flight observations of high energy photons from GRBs~\cite{Shao2010f, Zhang2015, Xu2016a, Xu2016,  Xu2018, Liu2018, ZHU2021136518} and active galactic nuclei~(AGNs)~\cite{Li2020}. These researches suggest \(\xi = E_\text{LV}^{-1}\), where \(E_\text{LV}=3.6\times 10^{17}~\text{GeV}\), see, e.g., Ref.~\cite{Xu2016}. While \(\xi_c^{-1}\simeq 4.5\times 10^{23}~\text{GeV}\), this value of \(\xi=E^{-1}_\text{LV}\) then falls into {\bfseries Region I}, and it indicates that CMB is more transparent to subluminal photons\footnote{This is true as long as \(E_\text{LV} < \xi_c^{-1}\).} than it is expected in special relativity. Even if one adopts a much stronger constraint on \(E_\text{LV}\) from other groups, such as \(E_\text{LV} \ge 0.58\times 10^{19}~\text{GeV}\) in Ref.~\cite{Magic}, the degree of freedom for \(E_\text{LV} \le \xi_c^{-1}\)~(which belongs to {\bfseries Region I}) is still large. Therefore we predict optical transparency for ultra-high energy cosmic photons with energies above \(k_0\simeq 411~\text{TeV}\), and any observation of such ultra-high energy cosmic photons from extragalactic sources can be considered as support for Lorentz invariance violation properties of photons.

    The aforementioned picture beyond special relativity might be attributed as unusual at first glance, but it is actually not odd since the threshold anomalies in LIV scenarios have already been studied extensively in the literature~(e.g.~\cite{Kifune:1999ex,Mattingly2003}) and our results are consistent with previous investigations.
    We make a brief review on some of these investigations and propose a novel strategy.
    The theory model of LIV induced threshold anomalies we adopt in this work is quite simple but with intuitive pictures, and in general one can consult Refs.~\cite{Mattingly2003,Jacobson2003} for rigorous formulations and systematic discussions of threshold anomalies in LIV\@.
    Besides these theoretical researches, LIV induced threshold anomaly phenomenology is studied intensively as well.
    Klu\'zniak~\cite{KLUZNIAK1999117}, Kifune~\cite{Kifune:1999ex} as well as Prothroe and Meyer~\cite{Protheroe:2000hp} analyzed the spectrums of two active galactic nuclei~(AGNs), Markarian 421 and Markarian 501, and the fact that these spectrums can smoothly extend to above the energy of 20~TeV brought the puzzle of an excess of ultra-high energy photons. Photons originating in distant sources such as AGNs with such high energies are not expected to be observable due to the attenuation by infrared background lights~\cite{Protheroe:2000hp}. Therefore Klu\'niak~\cite{KLUZNIAK1999117}, Kifune~\cite{Kifune:1999ex}, Prothroe and Meyer~\cite{Protheroe:2000hp} attributed this excess to LIV induced threshold anomalies allowing transparent phenomena similar to ours.
    Jacob and Piran~\cite{Jacob2008b} examined the absorbed spectra of extra-galactic gamma-ray sources such as blazars with LIV. They studied the threshold anomalies of TeV photons and EBL, and demonstrated a strategy that could test LIV by examining TeV blazar observations~\cite{Jacob2008b}.
    And based on this work~\cite{Jacob2008b}, Biteau and Williams~\cite{Biteau2015} studied TeV spectra of blazars in detail.
    Similarly, there are also intensive researches on LIV impacts on EBL opacity~\cite{Tavecchio2016,Franceschini2017, Abdalla2018}. Imprints from LIV induced EBL opacity modifications on spectra of especially blazars with TeV photons were taken into consideration in these researches, and possible observable phenomena were suggested.

    Most of the phenomenological works mentioned in the last paragraph focus on the properties of EBL opacity and the spectra of TeV photons from cosmological distance. We focus on CMB background photons and above or near PeV scale ultra-high energy photons. As a result in this work we propose an alternative and supplementary method of distinguishing LIV effect based on threshold anomalies besides examining the imprints of EBL opacity on the spectra of TeV photons from blazars or other cosmological objects. Our main start point is the transparency to ultra-high energy photons of the universe described in {\bfseries Case I} and {\bfseries Case II}, and it is of great importance since observations of this phenomenon can serve as definite signals of LIV\@. The strategy is as follows. Once we observe an above or near PeV photon, we check the possibility of it coming from a GRB~(or an AGN). The correlation between the photon and a potential GRB can be checked utilizing the criteria described in Refs.~\cite{Amelino-Camelia2016, Amelino-Camelia2017, Huang2018}. The direction of the potential source should be consistent with that of this photon and we adopt the following Gaussian to check this consistency~\cite{Amelino-Camelia2016, Amelino-Camelia2017, Huang2018}:
    \begin{equation}
        P(\gamma, \text{GRB})=\frac{1}{\sqrt{2\pi \sigma^2}}\exp(-\frac{\Delta\Phi^2}{2\sigma^2}),\label{directioncriterion}
    \end{equation}
    where \(\Delta\Phi \) is the angular separation between the GRB and the photon, and \(\sigma=\sqrt{\sigma^2_\text{GRB}+\sigma^2_\gamma}\) is the standard deviation of the angular uncertainties of the GRB and the photon measurements. A GRB and a photon are correlated if the angular separation between them are less than \(3\sigma{}\)~\cite{Huang2018}. After the angular correlation between the photon and a GRB is confirmed, we further examine the temporal correlation between them. According to Ref.~\cite{Jacob2008}, the LIV induced time of light flight delay of Eq.~(\ref{photondis}) is, for two photons with energies \(E_h\) and \(E_l\) respectively emitted from the source simultaneously,
    \begin{equation}
       \Delta t_\text{LV} = \xi \frac{E_h-E_l}{H_0}\int_0^z \frac{(1+z^\prime)dz^\prime}{\sqrt{\Omega_m(1+z^\prime)^3+\Omega_\Lambda}}, \label{tlv1}
    \end{equation}
    where \(z\) is the redshift of the GRB source, \([\Omega_m, \Omega_\Lambda]=[0.315^{+0.016}_{-0.017}, 0.685^{+0.017}_{-0.016}]\) are the cosmological constants and \(H_0=67.3\pm 1.2~ \text{km}~\text{s}^{-1}\text{Mpc}^{-1}\) is the present day Hubble constant~\cite{2014cgo}. We rewrite Eq.~(\ref{tlv1}) as
    \begin{equation}
        \Delta t_\text{LV} = (1+z) \xi K, \label{tlv2}
    \end{equation}
    where since \(E_h>>E_l\) we define
    \[
        K = \frac{E_h}{H_0(1+z)}\int_0^z \frac{(1+z^\prime)dz^\prime}{\sqrt{\Omega_m(1+z^\prime)^3+\Omega_\Lambda}}.
    \]
    As a result the observed time difference \(\Delta t_\text{obs}\) between these two photons is
    \begin{equation}
    \Delta t_\text{obs} = \Delta t_\text{LV} + (1+z)\Delta t_\text{in}, \label{tlv3}
    \end{equation}
    where \(\Delta t_\text{in}\) is the intrinsic time delay. Then we can check the temporal correlation between the photon and the GRB utilizing the following criteria~\cite{Amelino-Camelia2016, Amelino-Camelia2017, Huang2018}. From Eq.~(\ref{tlv3}) we have
    \begin{equation}
        \left\lvert \frac{\Delta t_\text{obs}}{1+z}-\Delta t_\text{in}\right\rvert=\left\lvert \xi K\right\rvert, \label{timecrit}
    \end{equation}
    and then we can utilize Eq.~(\ref{timecrit}) to check the temporal correlation. For example, we assume \(z=1\) and \(E_h=1 ~\text{PeV}\), and we take \(\Delta t_\text{in}\simeq -10~\text{s}\) and \(\xi^{-1}\simeq 3.6\times 10^{17}~\text{GeV}\)~\cite{Shao2010f, Xiao2009, Xiao2009a, Shao2010b, Zhang2015, Xu2016a, Xu2016,  Xu2018, Liu2018, ZHU2021136518, Li2020}. We have \(\left\lvert \xi K \right\rvert \simeq 11~\text{days}\). Therefore if the absolute value of the difference between the observation time of the photon and the tigger time of the GRB is less 11~days, there is a temporal correlation between the VHE photon and the GRB. Once both angular and temporal correlations are confirmed we think the VHE photon is from the GRB and further background analyses may be essential.

    It is worth noting that, recently, the LHAASO Collaboration reported the detection of ultra-high energy photons up to 1.4~PeV~\cite{Cao:2021}. It is an impressive result because LHAASO opens the window of observations of above-PeV photons which can shed light on several urgent and fundamental problems including Lorentz invariance violation~\cite{Li:2021pre2}. In special relativity, photons above 411~TeV are strongly absorbed by CMB and cannot propagate a long distance, and partly from this LHAASO considered these PeV photons as signals from PeVatrons~\footnote{These PeVatrons are located in the Milky Way and hence quite close to Earth so that PeV photons are still allowed to reach Earth.}.
    However, there still exists possibilities that at least some of the photons might come from extragalactic sources such as GRBs or AGNs once we consider the possibility of LIV\@. As we can see from {\bfseries Case I} and {\bfseries Case II}, the infrared background lights may be more transparent to ultra-high energy photons with LIV effect being considered, therefore PeV photons, which are expected to be absorbed strongly in special relativity, now can reach ground-based observatories. For these photons and more observations of PeV photons by LHAASO in the future, if we have solid evidence that some of them are from certain extragalactic sources, they can serve as convincing signals for LIV\@. We suggest the aforementioned strategy as a primary way to establish such correlations and more stringent methods are expected.

    In summary, we discuss the threshold anomalies of photon annihilation process \HepProcess{{\Pgg}{\Pgg}{\to}{\Pep}{\Pem}} due to Lorentz invariance violation, and reveal the novel phenomena of optical transparency, threshold reduction and reappearance of ultra-high energy photons in cosmic space.
    From a simple form of Lorentz violating dispersion relation supported by quantum gravity theories~\cite{Amelino-Camelia1997a, Ellis1999, Ellis2000,Ellis2008, Li2009, Li:2021pre, LI2021104380} as well as phenomenological analyses~\cite{Xu2016a,Xu2016,Xu2018,Li2020,ZHU2021136518}, we predict the optical transparency of ultra-high energy cosmic photons. Thus any observation of above threshold \(k_0\simeq 411~\text{TeV}\) photons from extragalactic sources can be considered as signals for new physics beyond special relativity.

    \section*{Acknowledgements}
        We thank Chengyi Li for helpful discussions. This work is supported by National Natural Science Foundation of China~(Grant No.~12075003).

\begin{onecolumn}

    \bibliography{ref-jhea}

\end{onecolumn}

\end{document}